\documentclass[%
 reprint,
 amsmath,amssymb,
 aps,
]{revtex4-1}

\usepackage{graphicx}% Include figure files
\usepackage{dcolumn}% Align table columns on decimal point
\usepackage{bm}% bold math
\usepackage{color}

\begin{document}

\preprint{APS/123-QED}

\title{Probing the Melting of a Two-dimensional Quantum Wigner Crystal\\via its Screening Efficiency}
% Force line breaks with \\

\author{H.\ Deng$^1$, L.N.\ Pfeiffer$^1$, K.W.\ West$^1$, K.W.\ Baldwin$^1$, L.W.\ Engel$^2$, and M.\ Shayegan$^1$}
%\author{H.\ Deng, \textit{et.al}}
\affiliation{$^1$Department of Electrical Engineering, Princeton University\\
$^2$National High Magnetic Field Laboratory, Tallahassee, Florida}

\date{\today}% It is always \today, today,

\begin{abstract}
One of the most fundamental and yet elusive collective phases of an interacting electron system
is the quantum Wigner crystal (WC), an ordered array of electrons expected to form
when the electrons' Coulomb repulsion energy eclipses their kinetic (Fermi) energy.
In low-disorder, two-dimensional (2D) electron systems,
the quantum WC is known to be favored at very low temperatures ($T$)
and small Landau level filling factors ($\nu$),
near the termination of the fractional quantum Hall states.
This WC phase exhibits an insulating behavior,
reflecting its pinning by the small but finite disorder potential.
An experimental determination of a $T$ vs $\nu$ phase diagram for the melting of the WC,
however, has proved to be challenging.
Here we use capacitance measurements to probe the 2D WC through its effective screening
as a function of $T$ and $\nu$.
We find that, as expected, the screening efficiency of the pinned WC is very poor at very low $T$
and improves at higher $T$ once the WC melts.
Surprisingly, however, rather than monotonically changing with increasing $T$,
the screening efficiency shows a well-defined maximum at a $T$
which is close to the previously-reported melting temperature of the WC.
Our experimental results suggest a new method to map out a $T$ vs $\nu$ phase diagram
of the magnetic-field-induced WC precisely.
\end{abstract}

\pacs{Valid PACS appear here}% PACS, the Physics and Astronomy
\maketitle

%Abstract

%Introduction
%	About WC: low T and high B, E_C > E_k, 2DES forms WC.
%	There are multiple experimental methods to probe WC.
%	In probing WC, the melting/phase transition is an important and interesting topic.
%	Previous experimental methods give their results, but with certain limits.
%		R_xx vs T ...
%		Nonlinear I-V/noise ...
%		RF resonance ...
%	In this work, we report probing WC via its screening efficiency.
%	Surprisingly, non-monotonic behavior of screening efficiency appears when T increases.
%	T_C at which 2DES screens most efficiently matches the reported melting T of WC.

In a Wigner crystal (WC) \cite{Wigner.PR.46.1002},
one of the earliest predicted many-body phases of an interacting electron system,
the dominance of electrons' Coulomb repulsion energy over their kinetic energy
forces them into a periodic array with long-range order.
In a two-dimensional electron system (2DES),
a quantum WC has long been expected to form
at low temperature ($T \lesssim 1$ K) and high magnetic field ($B$)
when the electrons occupy the lowest Landau level and their kinetic energy is quenched
\cite{Lozovik.JETP.22.11, Lam.PRB.30.473, Levesque.PRB.30.1056}.
There is also some experimental evidence, albeit often indirect,
for the formation of such a magnetic-field-induced, quantum WC
in very high-mobility (low-disorder) GaAs 2DESs near the Landau level filling factor $\nu = 1/5$
\cite{Andrei.PRL.60.2765, Jiang.PRL.65.633, Goldman.PRL.65.2189, Jiang.PRB.44.8107,
Williams.PRL.66.3285, Li.PRL.67.1630, Paalanen.PRB.45.13784, Li.SSC.95.619, MShayegan.WC.Review,
Pan.PRL.88.176802, Ye.PRL.89.176802, Ye.PRL.89.176802, Chen.NatPhys.2.245,
Tiemann.NatPhy.10.9.648, Hao.PRL.117.096601, Hao.PRB.98.081111}.
The main conclusion of these studies is that
the WC, being pinned by the ubiquitous residual disorder,
manifests in DC transport as an insulating phase with non-linear current-voltage (I-V) characteristics
\cite{Goldman.PRL.65.2189, Williams.PRL.66.3285, Jiang.PRB.44.8107, Li.PRL.67.1630},
and exhibits resonances in its high-frequency (microwave) AC transport
which strongly suggest collective motions of the electrons
\cite{Andrei.PRL.60.2765, Williams.PRL.66.3285, Paalanen.PRB.45.13784,
Li.SSC.95.619, MShayegan.WC.Review, Ye.PRL.89.176802, Chen.NatPhys.2.245}.
In a recent bilayer experiment, a high-density layer
hosting a composite fermion Fermi sea around $\nu = 1/2$ was used
to directly probe the microstructure of the WC forming in an adjacent, low-density layer
\cite{Hao.PRL.117.096601}.

A very fundamental property of the magnetic-field-induced WC
is its melting temperature vs filling factor phase diagram.
Probing the melting of such a WC, however, has been challenging.
Different experimental approaches have strived to determine the WC melting phase diagram,
but all the techniques face their own limitations.
One set of measurements showed ¡°kinks¡± in the Arrhenius plots of resistance vs $1/T$
which were used to extract a phase diagram \cite{Paalanen.PRB.45.13784};
however, such kinks were not reported by other groups
\cite{Jiang.PRL.65.633, Goldman.PRL.65.2189, Jiang.PRB.44.8107}.
The I-V measurements used the disappearance of the I-V non-linearity at high temperatures
to extract a melting temperature for the WC,
but the non-linearity often disappears very gradually and varies significantly from sample to sample
\cite{Goldman.PRL.65.2189, Williams.PRL.66.3285, Jiang.PRB.44.8107, Li.PRL.67.1630}.
The microwave resonance measurements also show broad resonance peaks at high $T$
and a rather gradual evolution with temperature,
making it difficult to pin the transition precisely \cite{Chen.NatPhys.2.245}.

Here, we probe the 2D WC through measuring, as a function of $T$ and $\nu$,
the capacitance between a top and a bottom gate that sandwich the 2DES.
Monitoring this capacitance provides a direct measure of the screening efficiency of the 2DES.
Similar measurements have demonstrated various properties of the 2DES
such as its compressibility \cite{Eisenstein.PRB.50.1760},
the incompressibility of quantum Hall states \cite{Zibrov.nphys.2018},
and a metal-insulator transition in relatively low-mobility samples \cite{Dultz.PRL.84.4689}.
Our data reveal an unexpected non-monotonic behavior for the screening efficiency of the 2DES
as it makes a transition from a pinned WC state at low $T$
to an interacting electron liquid at high $T$.
Most remarkably, the 2DES appears to be particularly good in screening at a $T$
which is close to the expected melting temperature of the WC.
This non-monotonic behavior is qualitatively different from the monotonic behaviors
we observe at other $\nu$ where the ground state of the 2DES is not a WC.
Associating the temperature at which the 2DES shows maximum screening with the melting temperature of the WC,
we determine a $T$ vs $\nu$ phase diagram which is tantalizingly similar
to those expected and reported for the WC.

%Experiment details
%	Sample structure, geometry, fabrication ...
%	Configurations of transport and I_P measurements. [Fig. 1 inset]

\begin{figure}[bp]
\includegraphics[width = 0.48\textwidth]{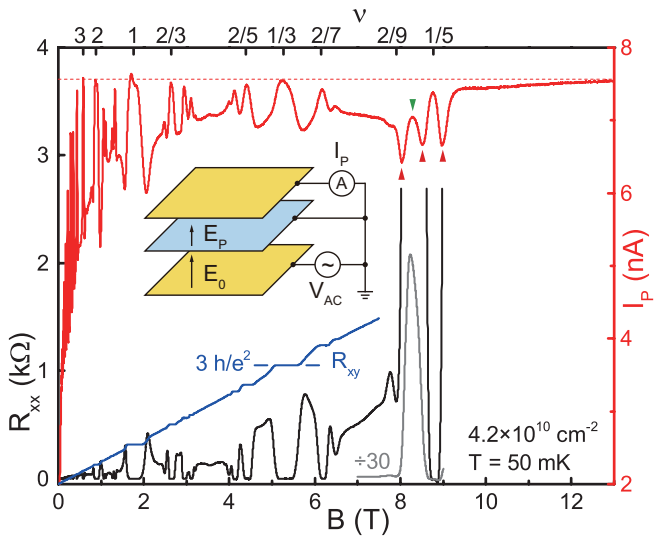}
\caption{
Overview of transport and penetration current results.
The inset shows a schematic of measurement configuration.
The yellow layers are the top and bottom gates,
and the blue layer is the 2DES.
An AC voltage ($V_{AC}$) applied to the bottom gate
generates the source electric field $E_{0}$ between this gate and the 2DES.
The penetration electric field $E_{P}$ reaches the top gate and induces $I_{P}$
which is measured by a lock-in amplifier in ammeter mode (A).
The black trace is the longitudinal magneto-resistance ($R_{xx}$);
the gray trace shows $R_{xx}$ reduced by a factor of 30 at high fields.
The blue trace is the Hall resistance ($R_{xy}$).
The red trace is the penetration current ($I_{P}$).
The $I_{P}$ maximum observed at the filling factor between 2/9 and 1/5,
where the reentrant WC is present, is marked by a green arrow.
The minima between the high-field WC and the FQHSs at $\nu = 2/9$ and 1/5
are indicated by the red triangles.
}
\end{figure}

Our sample is a modulation-doped, 70-nm-wide, GaAs quantum well (QW) grown via molecular beam epitaxy.
The 2DES has electron density $n = 4.2 \times 10^{10}$ cm$^{-2}$
with $\simeq 8.5 \times 10^{6}$ cm$^{2}/$Vs low-temperature mobility.
The sample has a van der Pauw ($\simeq 4 \times 4$ mm$^{2}$) geometry,
with six, alloyed In-Sn ohmic contacts made to the 2DES:
four on the corners of the sample, and two in the middle of two opposite edges.
The top and bottom gates are made from Ti-Au and In, respectively.
The distance between the QW and the top (bottom) gate is $\simeq 720$ nm ($\simeq 0.45$ mm).
The sample is cooled in a dilution refrigerator with a base temperature of $\simeq 37$ mK.
For in-plane, longitudinal ($R_{xx}$) and Hall ($R_{xy}$) transport measurements,
we use low-frequency lock-in technique at 7 Hz, while keeping the top and bottom gates grounded.
For measurements of the screening efficiency,
the configuration shown in Fig. 1 inset is used.
We apply a small (10 mV) AC voltage ($V_{AC}$) at 19 kHz to the bottom gate
and measure the current $I_{P}$ that penetrates to the top gate through the 2DES via a lock-in amplifier,
while all the contacts to the 2DES are grounded \cite{Footnote1}.
Large $I_{P}$ indicates low screening efficiency, and vice versa.
Note that, because of the small $V_{AC}$ amplitude and large distance between the QW and bottom gate,
the modulation of the 2DES density is negligible ($\simeq 10^{5}$ cm$^{-2}$).

%Results
%	General picture [Fig. 1]
%		Conventional transport measurement
%			R_xy, expected plateaus.
%			R_xx, expected minima
%				insulating states near \nu = 1/5.
%				WC and RWC.
%		I_P measurement
%			CF Fermi sea: compressible -> low I_P.
%			QHE: incompressible -> high I_P.
%			WC: "incompressible behavior" -> high I_P.
%			Note that I_P has minima around WC/RWC and FQHE.

Figure 1 provides an overview of our experimental results.
The in-plane transport traces, $R_{xx}$ and $R_{xy}$,
show the features expected for a high-mobility 2DES,
namely integer and fractional quantum Hall states (IQHS and FQHS):
$R_{xy}$ exhibits plateaus with expected values,
and $R_{xx}$ shows corresponding strong minima
at integer and fractional fillings as marked on the top axis.
Moreover, on the flanks of the well-developed $\nu = 1/5$ FQHS,
$R_{xx}$ shows highly resistive (insulating) states
which are attributed to the WC pinned by disorder
\cite{Jiang.PRL.65.633, Goldman.PRL.65.2189, Jiang.PRB.44.8107, Williams.PRL.66.3285,
Li.PRL.67.1630, Paalanen.PRB.45.13784, Li.SSC.95.619, MShayegan.WC.Review}.
For convenience, we denote the insulating state on the lower-$B$ side of $\nu = 1/5$ as the reentrant WC (RWC).

The measured $I_{P}$ (red trace in Fig. 1)
also reflects the rich phases of the 2DES as a function of $B$.
At zero magnetic field, the highly conductive 2DES
strongly screens the source electric field $E_{0}$ from the bottom gate,
resulting in a small $I_{P}$.
With increasing $B$, $I_{P}$ overall increases to a high level,
which is consistent with the general evolution of 2DES' bulk resistance \cite{Moon.PRL.79.4457}.
When the 2DES is in an IQHS or FQHS, $I_{P}$ shows a peak
because the 2DES bulk is in a gapped, incompressible state,
and therefore its screening efficiency is low.
In particular, for sufficiently strong QHSs,
the screening efficiency of the incompressible 2DES is negligible
so that the penetration electric field $E_{P}$
is essentially as strong as $E_{0}$,
leading to $I_{P}$ maxima at similar high values.
Indeed, at $\nu=3, 2, 1, 2/3$ and 1/3,
the measured $I_{P}$ has peaks with essentially the same height
(red, dashed horizontal line in Fig. 1).
Ideally, the $I_{P}$ maximum value is simply determined by the geometric capacitance
between the top and bottom gates as if the 2DES were not present.
The experimental value of $I_{P}$ maximum,
marked by the dashed red line in Fig. 1,
indeed agrees with our estimate ($7 \pm 2$ nA) based on the sample geometry,
namely the sample area and the distance between the top and bottom gates.
When the 2DES is in a compressible, liquid state,
e.g., at $\nu = 1/2$ and at $\nu$ between adjacent FQHSs,
$I_{P}$ is relatively low because now the 2DES bulk is compressible and conducting,
and therefore the 2DES screening efficiency is relatively high.

Most relevant to our study is the behavior of $I_P$ at very high magnetic fields
where the 2DES hosts the WC state:
$I_{P}$ has high value and tends to reach the same limit as seen for the strong IQHS/FQHSs.
This is consistent with the insulating behavior of the pinned WC
which should result in a low screening efficiency.
There is also a maximum in $I_{P}$ at a filling between 1/5 and 2/9, (green triangle in Fig. 1),
corresponding to the position of the RWC,
and there are maxima at $\nu = 1/5$ and 2/9 where incompressible FQHSs are present
(the maximum at $\nu=2/9$ is very weak because of the weakness of the 2/9 FQHS).
Between these maxima, there are three clear minima marked by red triangles in Fig. 1,
indicating the higher screening-efficiency states
separating the WC, RWC states and $\nu = 2/9$, 1/5 FQHSs.
In the remainder of the manuscript,
we carefully monitor $I_P$ as a function of $T$ and $\nu$
to elucidate the evolution of the various phases of the 2DES.
		
%	T dependence [Fig. 2]
%		Measure I_P vs B at different T, and full data set gives Fig. 2(a). [Fig. S1]
%		Measure I_P vs T for different 2DES phases at different B:
%			CF Fermi sea: relatively efficient screening, weak dependence on T.
%			FQHE: inefficient screening, weak negative dependence on T.
%			WC/RWC: non-monotonic behavior with T.
%		Discussion:
%			WC/RWC shows very different behavior than CF and QHE.
%			Exclude:
%				Non-ideal circuit: measurements at different freq show similar behavior. [Fig. S2 (a)]
%				Effect from 2DES bulk resistance: [Fig. 2(d,e)]
%					R_xx are monotonic with T.
%					R_xx are different when I_P reaches minimum at different \nu.
%			Possible explanation:
%				2DES is more compressible when it arranges itself close to an array. [Skinner]
%				Similar behavior observed before. [Eisenstein]
%					However, we are sure it's WC/RWC in our sample.
%				T_C for I-P minima suggests phase transition.
%					Compare to previous works: better defined T_C.

\begin{figure}[tbp]
\includegraphics[width = 0.48\textwidth]{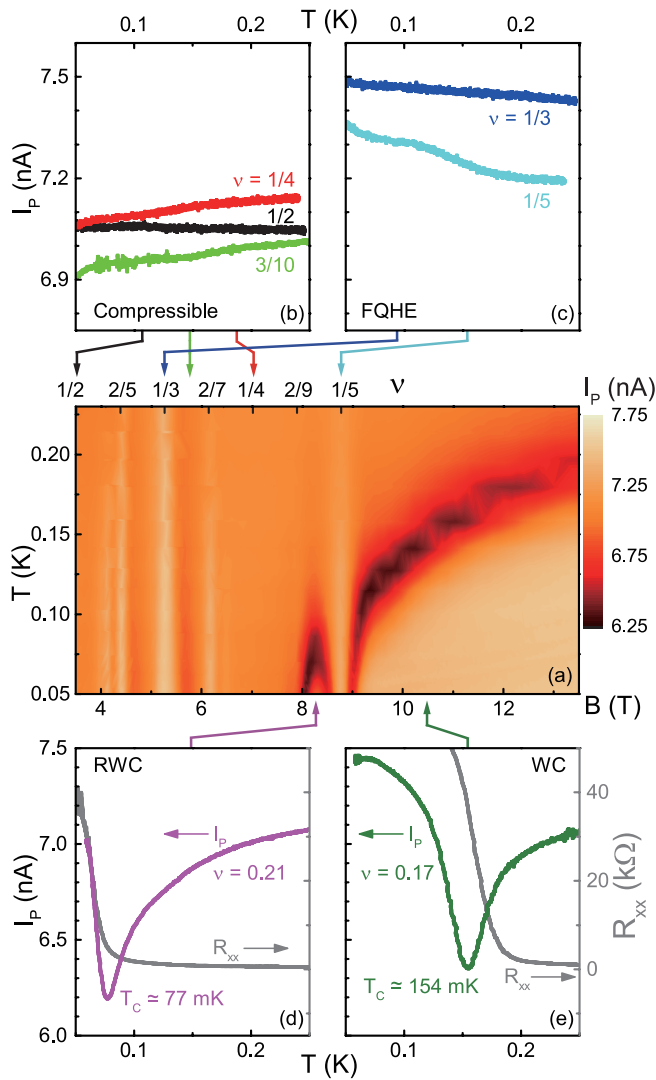}
\caption{
(a) Evolution of $I_{P}$ with magnetic field ($B$) and temperature ($T$).
(b)-(e) $T$-dependence of $I_{P}$ for various phases of the 2DES:
(b) compressible, electron liquid states at $\nu = 1/2, 1/4$ and $3/10$;
(c) incompressible FQHSs at $\nu = 1/3$ and $1/5$;
(d) reentrant Wigner crystal (RWC) phase at $\nu = 0.21$;
and (e) WC phase at $\nu = 0.17$.
In (d) and (e), the $T$-dependence of $R_{xx}$ (gray trace) at the same $\nu$ is also plotted.
(b) and (c) share the same $I_{P}$ scale on the left.
(d) and (e) have the same $I_{P}$ scale on the left, and $R_{xx}$ scale on the right (in gray).
The arrows on top of each panel indicate the $B$ positions for different phases.
}
\end{figure}

In Fig. 2(a) we present a color-density plot of $I_P$ as a function of $B$ and $T$.
This plot summarizes many $I_P$ vs $B$ traces taken at different $T$;
for typical traces in the high-$B$ range, see Fig. S1 in Supplemental Material \cite{Footnote.SM}.
The high-$I_{P}$ regimes (lighter color) locate the positions of FQHSs and WC/RWC,
while low-$I_{P}$ valleys (darker color) highlight the states separating these.
At the lowest $T$, there are three dark regions seen above $B = 8$ T ($\nu<2/9$);
these reflect the three $I_{P}$ minima marked by red triangles in Fig. 1.
In our experiments, we also independently measure $I_{P}$ vs $T$ at fixed $B$ for different 2DES phases,
and present the results in Figs. 2(b)-(e).
For the (compressible) Fermi liquid phases such as those at $\nu = 1/2, 3/10$ and $1/4$,
$I_{P}$ shows only a weak $T$ dependence and stays at a relatively low level [Fig. 2(b)].
This is consistent with the fact that these are metallic, conducting phases in the $T$ range of Fig. 2.
At FQHSs such as $\nu = 1/3$ and $1/5$, as seen in Fig. 2(c),
$I_{P}$ decreases monotonically from the high level at low $T$;
this is also expected as the quasi-particle excitations of these states
that are generated at higher $T$ are conducting and lead to screening.

However, the $T$ dependence of $I_{P}$ for the RWC and WC phases,
shown in Figs. 2(d) and (e), is qualitatively different from the monotonic behavior
seen for other phases of the 2DES.
With increasing temperature, $I_{P}$ decreases first,
reaches a minimum at a critical temperature ($T_{C}$) which depends on $\nu$,
and then increases and saturates at a value
which is lower than $I_{P}$ at base temperature.
This saturated $I_{P}$ value is almost the same in a large high-field range at high $T$,
implying the screening efficiency of the 2DES at high $T$
is nearly independent of filling and $T$ (also, see Fig. S2 in Supplemental Material).

The non-monotonic behavior of $I_{P}$ vs $T$ for the WC/RWC is surprising.
To ensure that it is not an artifact of our measurement circuit,
we repeated the measurements at multiple $V_{AC}$ frequencies,
covering over two orders of magnitude (10$^{2}$-10$^{4}$ Hz).
At different frequencies, $I_{P}$ vs $B$ traces show qualitatively similar behavior,
i.e., $I_{P}$ minima separate the FQHSs and the WC/RWC,
and $T$-dependence measurements show a similar non-monotonic behavior,
with $I_{P}$ minima at essentially the same $T_{C}$
(see Fig. S3 in Supplemental Material).
Also, the origin of $I_{P}$'s non-monotonic behavior
cannot be simply attributed to the changes in the 2DES bulk resistance.
In Figs. 2(d) and (e), we also plot the $T$ dependence of $R_{xx}$ (gray traces).
At both $\nu = 0.21$ [Fig. 2(d)] and $\nu = 0.17$ [Fig. 2(e)],
$R_{xx}$ shows a monotonic dependence on $T$ as expected \cite{Jiang.PRL.65.633},
which is hard to link with the non-monotonic behavior of $I_{P}$.
Moreover, at different $\nu$, when $I_{P}$ reaches the minimum at $T_{C}$,
$R_{xx}$ has significantly different values [$\sim 4$ and 40 k$\Omega$ in Figs. 2(d) and (e), respectively],
indicating that the $I_{P}$ minimum is not associated with a certain value of $R_{xx}$.

We suspect that the $I_{P}$ minimum, i.e., the max screening efficiency,
signals a phase transition in the WC/RWC.
Theory suggests that an interacting 2DES is more compressible
when it arranges itself close to an ordered array which has strong positional correlation \cite{Skinner.PRB.88.155417}.
Indeed, in previous studies of negative compressibility in a bilayer 2DES \cite{Eisenstein.PRB.50.1760},
the 2DES shows enhanced screening efficiency (even over-screening)
when $\nu$ decreases (by lowering the 2DES density at fixed $B$),
and suddenly loses its screening ability at very small $\nu$ ($\lesssim 0.05$).
In Ref. \cite{Eisenstein.PRB.50.1760}, this sudden transition in screening efficiency
was attributed to the localization of electrons in the random disorder potential.

This interpretation might be reasonable for the samples of Ref. \cite{Eisenstein.PRB.50.1760},
which had relatively lower quality
(mobility of $\simeq 2.1 \times 10^{6}$ cm$^{2}/$Vs at $n = 7.5 \times 10^{10}$ cm$^{-2}$),
and considering that the transition in screening efficiency happened
when the 2DES was depleted to a very low $n$ ($< 1 \times 10^{10}$ cm$^{-2}$).
In contrast, $n$ is much higher ($4.2 \times 10^{10}$ cm$^{-2}$) in our entire measurement range,
and the sample's high quality is evinced by the very large mobility ($\simeq 8.5 \times 10^{6}$ cm$^{2}/$Vs),
as well as the rich sequence of FQHSs, and especially the well-developed $\nu = 1/5$ FQHS.
These facts strongly suggest that the high-$I_{P}$, low-screening states at low fillings
in our sample reflect the formation of (pinned) collective WC and RWC states
rather than the onset of single-electron localization by strong disorder.
We therefore surmise that the positions of $I_{P}$ minima in Fig. 2(a)
might imply a phase transition in the WC/RWC.

%	Extract T_C vs \nu from Fig. 2(a). [Fig. 3]
%		RWC and WC are separated by FQHE.
%		RWC is flanked by FQHEs.
%		Data from I_P vs T at fix \nu fall on the same trend.
%		Discussion:
%			Shows qualitatively similar phase diagram to previous works.
%			If T_C is melting T, generally lower:
%				Reason is not clear.
%				Wider QW, lower melting T. [Lloyd]
%			Data from higher n:
%				Qualitatively same behavior,
%				Higher T_C: stronger e-e interaction.

\begin{figure}[tbp]
\includegraphics[width = 0.48\textwidth]{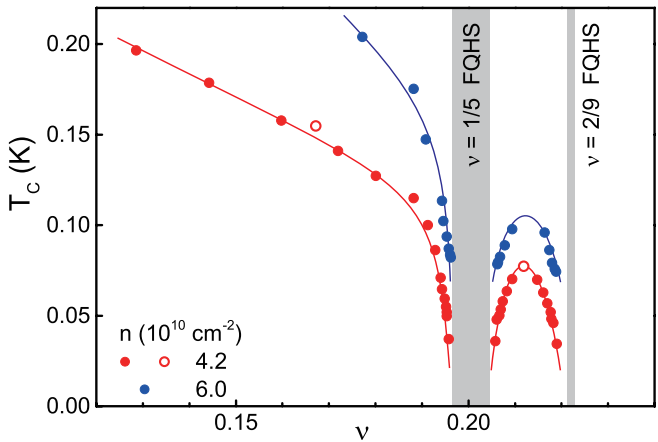}
\caption{
The measured critical temperature ($T_{C}$) vs $\nu$.
The red and blue circles are the data measured at $n = 4.2$ and $6.0 \times 10^{10}$ cm$^{-2}$, respectively.
The data represented by solid circles are measured by sweeping $B$ at fixed $T$ [Fig. 2(a)],
while the empty circles are measured by sweeping $T$ at fixed $B$ [Figs. 2(d)-(e)].
The gray zones indicate the regime of nearby FQHSs.
The curves connecting data points serve as guides to the eye.
}
\end{figure}

From Fig. 2(a), we extract the positions of $I_{P}$ minima
and plot the data in Fig. 3 as solid red circles;
the open red circles are from the $I_{P}$ vs $T$ data of Figs. 2(d) and 2(e).
The RWC "dome" is flanked by $\nu = 1/5$ and 2/9 FQHSs (gray zones),
and the $\nu = 1/5$ FQHS separates the RWC and WC.
The $\nu$ dependence of $T_{C}$ in Fig. 3
is qualitatively consistent with the WC-liquid phase diagrams reported previously
\cite{Goldman.PRL.65.2189, Williams.PRL.66.3285, Paalanen.PRB.45.13784, Chen.NatPhys.2.245}.
Associating $T_{C}$ with the melting temperature of the WC/RWC,
the well-pronounced and relatively sharp $I_{P}$ minima in our measurements
allow us to unveil rich details of the melting,
especially the clear "dome"-shaped boundary of the RWC.
We note that our data imply somewhat lower melting temperatures
compared to those reported in Ref. \cite{Chen.NatPhys.2.245}
for 2DES samples with quality comparable to ours.
A possible reason for this discrepancy might be that,
unlike the GaAs/AlGaAs hetero-structures and narrow QWs used in Ref. \cite{Chen.NatPhys.2.245},
the 2DES in our wide (70 nm) QW sample has a larger layer thickness.
This larger thickness can soften the Coulomb interaction
between electrons and lower the WC melting temperature.

We also repeated similar measurements and analysis of $I_{P}$
at a higher 2DES density of $n=6.0 \times 10^{10}$ cm$^{-2}$,
attained by applying DC voltage biases to the top and bottom gates.
The values of gate voltages were set carefully
to keep the charge distribution in the QW symmetric.
The $T_{C}$ vs $\nu$ data for the higher density are plotted as blue circles in Fig. 3.
The data are qualitatively similar to the data measured at lower $n$,
but exhibit a larger $T_{C}$ in general.
The overall higher $T_{C}$
is consistent with the stronger electron-electron interaction at higher $n$
which should help stabilize the WC
and therefore increase the melting temperature at a given filling.

%Conclusion
%	Clear, well defined features (I_P minima) between WC/RWC and liquid phases.
%	If T_C can be associated with melting T, it would be a good way to map out phase diagram.

In conclusion, via measurements of the penetrating current $I_{P}$ through a 2DES,
we probe its screening efficiency.
The data show very different behaviors at different filling factors
as the 2DES goes through its many-body states.
In particular, the WC and RWC states exhibit a high $I_P$, revealing low screening efficiency.
Surprisingly, $I_{P}$ shows well-defined minima
as a function of either filling or temperature for the WC and RWC states,
and the positions of the minima are consistent with the melting of these states.
If the $T_{C}$ of $I_{P}$ minima could indeed be associated with the WC melting temperature,
our data demonstrate that the measurements of screening efficiency
provide a prime technique to map out the phase diagram of the magnetic-field-induced WC precisely.
Regardless of its possible association with the WC melting,
the non-monotonic behavior for the screening efficiency we observe is novel,
and begs theoretical explanation.

%Acknowledgement
\begin{acknowledgments}
We thank R. N. Bhatt, M. Dykman, D. Huse, J. K. Jain, S. Kivelson and S. Sondhi for helpful discussions.
We acknowledge the National Science Foundation (Grant DMR 1709076) for measurements,
and the Gordon and Betty Moore Foundation
(Grant GBMF4420), the Department of Energy Basic Energy Sciences (Grant
DE-FG02-00-ER45841), and the National Science Foundation (Grants MRSEC DMR 1420541 and
ECCS 1508925) for sample fabrication.
L. W. E. is supported by the Department of Energy (Grant DE-FG02-05-ER46212).
\end{acknowledgments}

\bibliographystyle{h-physrev-title}
%\bibliography{WCC_cite}

\begin{thebibliography}{10}
\bibitem{Footnote1}
In all the plots presented in our manuscript, we show the magnitude of the
  measured current $I_P$. The out-of-phase component of $I_P$ is typically much
  larger than its in-phase component, and both components show qualitatively
  similar behaviors as a function of filling and/or temperature.
\end{thebibliography}

\end{document}